\documentclass[10pt,a4paper,english,conference]{IEEEtran}
\usepackage[T1]{fontenc}
\usepackage{verbatim}
\usepackage{float}
\usepackage{mathrsfs}
\usepackage{amsmath}
\usepackage{amsthm}
\usepackage{amssymb}
\usepackage{graphicx}
\usepackage{setspace}

\makeatletter

\pdfpageheight\paperheight
\pdfpagewidth\paperwidth

\floatstyle{ruled}
\newfloat{algorithm}{tbp}{loa}
\providecommand{\algorithmname}{Algorithm}
\floatname{algorithm}{\protect\algorithmname}

\theoremstyle{plain}
\newtheorem{thm}{\protect\theoremname}
\theoremstyle{plain}
\newtheorem{prop}[thm]{\protect\propositionname}

\usepackage{subfigure}
\usepackage{epstopdf}
\usepackage{balance}
\usepackage{cite}
\usepackage{citesort}
\IEEEoverridecommandlockouts

\makeatother

\usepackage{babel}
\providecommand{\propositionname}{Proposition}
\providecommand{\theoremname}{Theorem}

\begin{document}
% paper title

\title{Study on the Idle Mode Capability with \\
LoS and NLoS Transmissions% <-this % stops a space
}

\author{\noindent {\normalsize{}Ming Ding$^{\ddagger}$, David L$\acute{\textrm{o}}$pez
P$\acute{\textrm{e}}$rez$^{\dagger}$}\emph{\normalsize{}, }{\normalsize{}Guoqiang
Mao$^{\nparallel}{}^{\ddagger}{}^{\sqcap}{}^{\mathsection}{}^{\clubsuit}$}\thanks{$^{\clubsuit}$Guoqiang Mao's research is supported by the Australian
Research Council (ARC) Discovery projects DP110100538 and DP120102030
and the Chinese National Science Foundation project 61428102.}\emph{\normalsize{}, }{\normalsize{}Zihuai Lin}\textit{\normalsize{}$^{\mathparagraph}$}\emph{\normalsize{}}\\
\textit{\small{}$^{\ddagger}$Data61, Australia }{\small{}\{Ming.Ding@data61.csiro.au\}}\textit{\small{}}\\
\textit{\small{}$^{\dagger}$Nokia Bell Labs, Ireland }{\small{}\{david.lopez-perez@nokia.com\}}\textit{\small{}}\\
\textit{\small{}$^{\nparallel}$School of Computing and Communication,
University of Technology Sydney, Australia}\\
\textit{\small{}$^{\sqcap}$School of Electronic Information \& Communications,
Huazhong University of Science \& Technology, Wuhan, China}\\
\textit{\small{}$^{\mathsection}$School of Information and Communication
Engineering, Beijing University of Posts and Telecommunications, Beijing,
China}\\
\textit{\small{}$^{\mathparagraph}$The University of Sydney, Australia}% <-this % stops a space
}
\maketitle
\begin{abstract}
In this paper, we study the impact of the base station (BS) idle mode
capability (IMC) on the network performance in dense small cell networks
(SCNs). Different from existing works, we consider a sophisticated
path loss model incorporating both line-of-sight (LoS) and non-line-of-sight
(NLoS) transmissions. Analytical results are obtained for the coverage
probability and the area spectral efficiency (ASE) performance for
SCNs with IMCs at the BSs. The upper bound, the lower bound and the
approximate expression of the activated BS density are also derived.
The performance impact of the IMC is shown to be significant. As the
BS density surpasses the UE density, thus creating a surplus of BSs,
the coverage probability will continuously increase toward one. For
the practical regime of the BS density, the results derived from our
analysis are distinctively different from existing results, and thus
shed new light on the deployment and the operation of future dense
SCNs.
\footnote{1536-1276 © 2015 IEEE. Personal use is permitted, but republication/redistribution requires IEEE permission. Please find the final version in IEEE from the link: http://ieeexplore.ieee.org/document/7842302/. Digital Object Identifier: 10.1109/GLOCOM.2016.7842302}
\end{abstract}

\begin{comment}
Keywords: stochastic geometry, homogeneous Poisson point process (HPPP),
Line-of-sight (LoS), Non-line-of-sight (NLoS), dense small cell networks
(SCNs), coverage probability, area spectral efficiency (ASE). Title
for journal paper: Performance Analysis of Dense Cellular Networks
with the Idle Mode Capability
\end{comment}

\section{Introduction\label{sec:Introduction}}

\begin{comment}
Placeholder
\end{comment}
\begin{comment}
Driven by a new generation of wireless user equipment (UE) and the
proliferation of bandwidth-intensive applications, mobile data traffic
and network load are increasing in an exponential manner, and are
straining current cellular networks to a breaking point~\cite{Report_CISCO}.
\end{comment}

Dense small cell networks (SCNs)%
\begin{comment}
, comprised of remote radio heads, metrocells, picocells, femtocells,
relay nodes, etc.,
\end{comment}
{} have attracted much attention as one of the most promising approaches
to rapidly increase network capacity and meet the ever-increasing
capacity demands~\cite{Report_CISCO}. Indeed, the orthogonal deployment\footnote{The orthogonal deployment means that small cells and macrocells operate
on different frequency spectrum, i.e., Small Cell Scenario \#2a defined
in~\cite{TR36.872}.} of dense SCNs within the existing macrocell networks~\cite{TR36.872}
\begin{comment}
i.e., small cells and macrocells operating on different frequency
spectrum (Small Cell Scenario \#2a~\cite{TR36.872}),
\end{comment}
has been selected as the workhorse for capacity enhancement in the
3rd Generation Partnership Project (3GPP) 4th-generation (4G) and
the 5th-generation (5G) networks. This is due to its large spectrum
reuse and its easy management~\cite{Tutor_smallcell}; the latter
one arising from its low interaction with the macrocell tier, e.g.,
no inter-tier interference. In this paper, the focus is on the analysis
of these dense SCNs with an orthogonal deployment in the existing
macrocell networks.

\begin{comment}
Dense SCNs can achieve a high spatial spectrum reuse by creating a
large number of small cells through the network densification, which
in turn can significantly enhance network capacity through cell splitting
gains~\cite{Tutor_smallcell}. Note that the orthogonal deployment
of dense SCNs within the existing macrocell network, i.e., small cells
and macrocells operate on different frequency spectrum (Small Cell
Scenario \#2a defined in~\cite{TR36.872}), is gaining much momentum
in the design of the 4th-generation (4G) Long Term Evolution (LTE)
networks by the 3rd Generation Partnership Project (3GPP), and is
envisaged to be the workhorse for capacity enhancement in the 5th-generation
(5G) networks due to its large performance gains and its easy deployment,
arising from its low interaction with the macrocell tier, e.g., no
inter-tier interference~\cite{Tutor_smallcell}. In this paper, our
focus is on these dense SCNs with an orthogonal deployment with the
macrocell network.
\end{comment}
\begin{comment}
Indeed, SCNs have attracted much momentum in the wireless communications
industry and research community~\cite{Tutor_smallcell}, and have
also gained the attention of standardization bodies such as the 3rd
Generation Partnership Project (3GPP) in the design of Long Term Evolution
(LTE) networks~\cite{TR36.872}.
\end{comment}

Before 2015, the common understanding on dense SCNs was that the density
of base stations (BSs) would not affect the per-BS coverage probability
performance in interference-limited fully-loaded wireless networks~\cite{Jeff2011},
where the coverage probability is defined as the probability that
the signal-to-interference-plus-noise ratio (SINR) of a typical user
equipment (UE) is above a SINR threshold $\gamma$.%
\begin{comment}
, i.e., all BSs are activated considering an infinite user equipment
(UE) density.
\end{comment}
{} Consequently, the area spectral efficiency (ASE) performance in $\textrm{bps/Hz/km}^{2}$
would scale linearly with the network densification~\cite{Jeff2011}.
\begin{comment}
The implication of such conclusion is huge: \emph{The BS density does
NOT matter}, since the increase in the interference power caused by
a denser network would be exactly compensated by the increase in the
signal power due to the reduced distance between transmitters and
receivers.
\end{comment}
\begin{comment}
we do need to worry about the \emph{quantity} of interferers in dense
SCNs because the useful signal always has a superior \emph{quality}
compared with the interference. Thus,
\end{comment}
The intuition of such conclusion is that the increase in the interference
power caused by a denser network would be exactly compensated by the
increase in the signal power due to the reduced distance between transmitters
and receivers. Fig.~\ref{fig:comp_p_cov_4Gto5G} shows this theoretical
coverage probability behavior predicted in~\cite{Jeff2011}. %
\begin{comment}
, which also exhibits the typical BS density regions for various generations
of telecommunication systems.
\end{comment}
However, it is important to note that such conclusion was obtained
with considerable simplifications on the UE deployment and propagation
environment, e.g., all BSs were activated considering an infinite
UE density. Moreover, a single-slope path loss model was used, which
should be placed under scrutiny when evaluating practical dense SCNs,
since they are fundamentally different from sparse ones%
\begin{comment}
 in various aspects
\end{comment}
~\cite{Tutor_smallcell}.

\noindent \begin{center}
\begin{figure}
\noindent \begin{centering}
\vspace{-0.2cm}
\includegraphics[width=7cm]{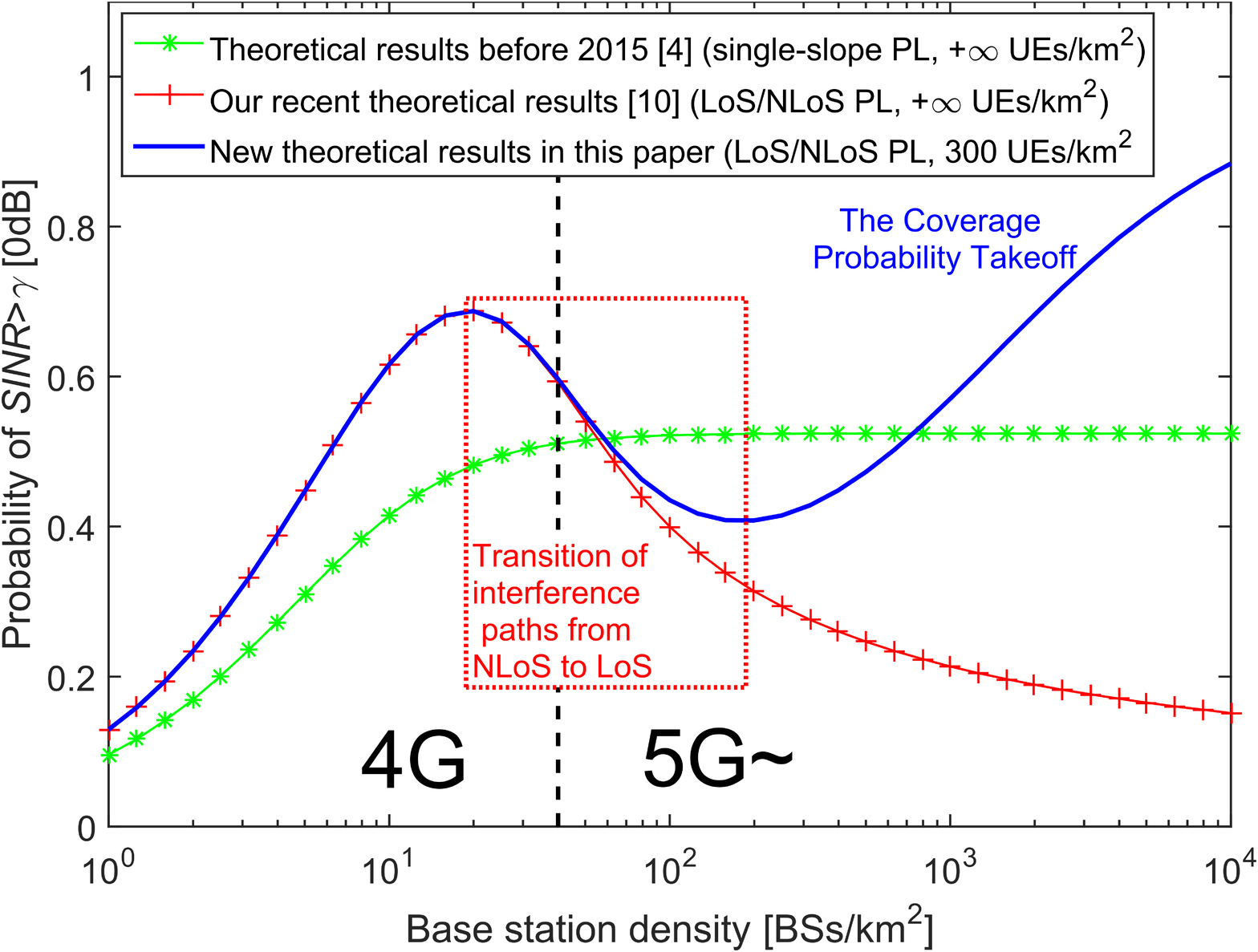}\renewcommand{\figurename}{Fig.}\caption{\label{fig:comp_p_cov_4Gto5G}Theoretical comparison of the coverage
probability performance when the SINR threshold $\gamma=0$\,dB.
Note that all the results are obtained using practical 3GPP channel
models~\cite{TR36.828,SCM_pathloss_model}, which will be introduced
in details later. Moreover, the BS density regions for the 4G and
the 5G networks have been illustrated in the figure, considering that
the typical BS density of the 4G SCNs is in the order of tens of $\textrm{BSs/km}^{2}$~\cite{TR36.872,Tutor_smallcell}. }
\par\end{centering}
\vspace{-0.5cm}
\end{figure}
\par\end{center}

\vspace{-0.9cm}

A few noteworthy studies have been carried out in the last year %
\begin{comment}
 in this regard, revisiting
\end{comment}
to revisit the network performance analysis for dense SCNs under %
\begin{comment}
and embracing
\end{comment}
more practical propagation assumptions. In~\cite{related_work_Jeff},
the authors considered a multi-slope piece-wise path loss function,
while in~\cite{Related_work_Health}, the authors investigated line-of-sight
(LoS) and non-line-of-sight (NLoS) transmission as a probabilistic
event for a millimeter wave communication scenario. %
\begin{comment}
The most important finding in these two works was that the per-BS
coverage probability performance starts to decrease when the BS density
is sufficiently large. Fortunately, such decrease of coverage probability
did not change the monotonic increase of the ASE as the BS density
increases.
\end{comment}
In our very recent work~\cite{our_GC_paper_2015_HPPP,our_work_TWC2016},
we took a step further and generalized these works by considering
both piece-wise path loss functions and probabilistic LoS and NLoS
transmissions. Our new finding was not only quantitatively but also
qualitatively different from the results in~\cite{Jeff2011,related_work_Jeff,Related_work_Health}.
In more detail, our analysis demonstrated that %
\begin{comment}
the coverage probability performance will initially increase with
the increase of the BS density. But
\end{comment}
when the BS density is larger than a threshold $\lambda^{*}$, the
coverage probability performance will decrease as the SCN becomes
denser, which in turn may make the ASE suffer from a slow growth or
even a\emph{ decrease} on the journey from 4G to 5G. The intuition
behind this result is that the interference power increases faster
than the signal power due to the transition of a large number of interference
paths from NLoS to LoS with the network densification. Our analysis
also demonstrated that the decrease of the coverage probability gradually
slows down as the SCN becomes ultra-dense. This is because both the
interference power and the signal power are LoS dominated and thus
statistically stable. Fig.~\ref{fig:comp_p_cov_4Gto5G} shows this
new theoretical coverage probability result, where $\lambda^{*}$
is around 20\,$\textrm{BSs/km}^{2}$.%
\begin{comment}
The decrease of the coverage probability gradually slows down as the
BS density goes ultra-high, because both the signal power and the
interference power are LoS dominated and thus statistically stable.
Fig.~\ref{fig:comp_p_cov_4Gto5G} shows these new theoretical results
in terms of the coverage probability, where such BS density threshold
$\lambda^{*}$ is around 20\,$\textrm{BSs/km}^{2}$. The intuition
of our conclusion is that the interference power increases faster
than the signal power due to the transition of a large number of interference
paths from NLoS to LoS with the network densification.
\end{comment}
\begin{comment}
The implication is profound:
\end{comment}
\begin{comment}
we need to be careful about the \emph{quality} of interferers because
many of them may transit from NLoS to LoS, thus overwhelming the signal
link at certain BS density ranges. Thus,
\end{comment}
\emph{}%

Fortunately, the UE density is finite in practice, and thus a large
number of SCN BSs could be switched off in dense SCNs, if there is
no active UE within their coverage areas, which mitigates unnecessary
inter-cell interference and reduces energy consumption~\cite{dynOnOff_Huang2012,Luo2014_dynOnOff,Li2014_energyEff}.
In more detail, by dynamically turning off idle BSs, the interference
suffered by UEs from always-on channels, e.g., synchronization and
broadcast channels, and data channels can be reduced, thus improving
UEs' coverage probability. Besides, the energy efficiency of the SCNs
can be significantly enhanced because (i) BSs without any active UE
can be put into a zero/low-power idle mode until a UE becomes active
in its coverage area, and (ii) activated BSs usually enjoy high-SINR
links, i.e., energy-efficient links, with their associated UEs thanks
to the BS selection from a surplus of BSs.%
\begin{comment}
Such BS capability will be referred to as the idle mode capability
(IMC) hereafter.
\end{comment}
{} Such capability of idle mode at the BSs is referred to as the idle
mode capability (IMC) hereafter.

In this paper, we investigate for the first time the impact of the
IMC on the coverage probability performance. Our new theoretical results
with a UE density of $300\thinspace\textrm{UEs/km}^{2}$, a typical
UE density in 5G~\cite{Tutor_smallcell}, are compared with the existing
results~\cite{Jeff2011,our_work_TWC2016} in Fig.~\ref{fig:comp_p_cov_4Gto5G}.
The performance impact of the IMC is shown to be significant: as the
BS density surpasses the UE density, thus creating a surplus of BSs%
\begin{comment}
 in dense SCNs
\end{comment}
, the coverage probability will continuously increase toward one,
addressing the issue caused by the NLoS to LoS transition of interfering
paths. Such performance behavior is referred to as\emph{ the Coverage
Probability Takeoff} hereafter. The intuition behind \emph{the Coverage
Probability Takeoff} is that the interference power will remain constant
with the network densification due to the IMC at the BSs, while the
signal power will continuously grow due to the BS selection and the
shorter BS-to-UE distance, thus permitting stronger serving BS links.

\begin{comment}
Compared with our previous works~\cite{our_GC_paper_2015_HPPP,our_work_TWC2016},
our new discovery has been achieved by changing nothing but adopting
one 3GPP assumption that the BS antenna height and the UE antenna
height are set to 10\,m and 1.5\,m, respectively~\cite{TR36.814}.
\end{comment}

The main contributions of this paper are as follows:
\begin{itemize}
\item Analytical results are obtained for the coverage probability and the
ASE performance for SCNs with IMC at the BSs using a general path
loss model incorporating both LoS and NLoS transmissions. Note that
the existing works on the IMC only treat single-slope path loss models
where UEs are always associated with the nearest BS~\cite{dynOnOff_Huang2012,Li2014_energyEff},
while our work considers more practical path loss models with LoS
and NLoS transmissions where UEs may connect to a farther BS with
a LoS path.%
\begin{comment}
Using the above results assuming a general path loss model, numerically
tractable integral-form expressions for the coverage probability and
the ASE are further obtained for a 3GPP path loss model with a \emph{linear}
LoS probability function. Our analysis can be readily extended to
deal with complicated path loss models by approximating the corresponding
LoS probability function as a piece-wise \emph{linear} function.
\end{comment}
\item The upper bound, the lower bound and the approximate expression of
the activated BS density are derived for the SCNs with IMCs considering
 practical path loss models with LoS and NLoS transmissions.%
\begin{comment}
Our theoretical analysis reveals an important finding, i.e., when
the density of small cells is larger than a threshold, the network
coverage probability will decrease as small cells become denser, which
in turn may make the ASE suffer from a slow growth or even a decrease.
The ASE will grow almost linearly as the base station (BS) density
increases above another larger threshold. This finding that the ASE
performance may suffer from a slow growth or even a decrease as the
BS density increases, is not only quantitatively but also qualitatively
different from previous study results with a simplistic path loss
model that does not differentiate LoS and NLoS transmissions. Although
our conclusion is made from the investigated set of parameters recommended
by the 3GPP for SCNs, which sheds valuable insights on the design
and deployment of practical SCNs, it is of significant interest to
further study the generality of our conclusion in other network models
and with other parameter sets.
\end{comment}
\end{itemize}

\section{System Model\label{sec:System-Model}}

\begin{comment}
Placeholder
\end{comment}

We consider a downlink (DL) cellular network with BSs deployed on
a plane according to a homogeneous Poisson point process (HPPP) $\Phi$
of intensity $\lambda$ $\textrm{BSs/km}^{2}$%
\begin{comment}
(i.e., $\textrm{cells/km}^{2}$)
\end{comment}
. Active UEs are Poisson distributed in the considered network with
an intensity of $\rho$ $\textrm{UEs/km}^{2}$. Here, we only consider
active UEs in the network because non-active UEs do not trigger data
transmission, and thus they are ignored in our analysis.

In our previous works~\cite{our_GC_paper_2015_HPPP,our_work_TWC2016}
and other related works~\cite{related_work_Jeff,Related_work_Health},
$\rho$ was assumed to be sufficiently larger than $\lambda$ so that
each BS has at least one associated UE in its coverage. In this work,
we impose no such constraint on $\rho$, and hence a BS with the IMC
will be switched off if there is no UE connected to it, which reduces
interference to neighboring UEs as well as energy consumption. Since
UEs are randomly and uniformly distributed in the network, we adopt
a common assumption that the activated BSs also follow an HPPP distribution
$\tilde{\Phi}$~\cite{dynOnOff_Huang2012}, the intensity of which
is denoted by $\tilde{\lambda}$ $\textrm{BSs/km}^{2}$. Note that
$0\leq\tilde{\lambda}\leq\lambda$ and a larger $\rho$ requires more
BSs with a larger $\tilde{\lambda}$ to serve the active UEs.

Following~\cite{our_GC_paper_2015_HPPP,our_work_TWC2016}, we adopt
a very general and practical path loss model, in which the path loss
$\zeta\left(r\right)$ associated with distance $r$ is segmented
into $N$ pieces written as%
\begin{comment}
\begin{singlespace}
\noindent
\[
\zeta\left(w\right)=\begin{cases}
\zeta_{1}\left(w\right)=\begin{cases}
\begin{array}{l}
\zeta_{1}^{\textrm{L}}\left(w\right),\\
\zeta_{1}^{\textrm{NL}}\left(w\right),
\end{array} & \hspace{-0.3cm}\begin{array}{l}
\textrm{with probability }\textrm{Pr}_{1}^{\textrm{L}}\left(w\right)\\
\textrm{with probability }\left(1-\textrm{Pr}_{1}^{\textrm{L}}\left(w\right)\right)
\end{array}\end{cases}\hspace{-0.3cm}, & \hspace{-0.3cm}\textrm{when }0\leq w\leq d_{1}\\
\zeta_{2}\left(w\right)=\begin{cases}
\begin{array}{l}
\zeta_{2}^{\textrm{L}}\left(w\right),\\
\zeta_{2}^{\textrm{NL}}\left(w\right),
\end{array} & \hspace{-0.3cm}\begin{array}{l}
\textrm{with probability }\textrm{Pr}_{2}^{\textrm{L}}\left(w\right)\\
\textrm{with probability }\left(1-\textrm{Pr}_{2}^{\textrm{L}}\left(w\right)\right)
\end{array}\end{cases}\hspace{-0.3cm}, & \hspace{-0.3cm}\textrm{when }d_{1}<w\leq d_{2}\\
\vdots & \vdots\\
\zeta_{N}\left(w\right)=\begin{cases}
\begin{array}{l}
\zeta_{N}^{\textrm{L}}\left(w\right),\\
\zeta_{N}^{\textrm{NL}}\left(w\right),
\end{array} & \hspace{-0.3cm}\begin{array}{l}
\textrm{with probability }\textrm{Pr}_{N}^{\textrm{L}}\left(w\right)\\
\textrm{with probability }\left(1-\textrm{Pr}_{N}^{\textrm{L}}\left(w\right)\right)
\end{array}\end{cases}\hspace{-0.3cm}, & \hspace{-0.3cm}\textrm{when }w>d_{N-1}
\end{cases}.
\]
\end{singlespace}
\end{comment}

\begin{singlespace}
\noindent
\begin{equation}
\zeta\left(r\right)=\begin{cases}
\zeta_{1}\left(r\right), & \textrm{when }0\leq r\leq d_{1}\\
\zeta_{2}\left(r\right), & \textrm{when }d_{1}<r\leq d_{2}\\
\vdots & \vdots\\
\zeta_{N}\left(r\right), & \textrm{when }r>d_{N-1}
\end{cases},\label{eq:prop_PL_model}
\end{equation}

\end{singlespace}

\noindent where each piece $\zeta_{n}\left(r\right),n\in\left\{ 1,2,\ldots,N\right\} $
is modeled as

\noindent
\begin{equation}
\zeta_{n}\left(r\right)\hspace{-0.1cm}=\hspace{-0.1cm}\begin{cases}
\hspace{-0.2cm}\begin{array}{l}
\zeta_{n}^{\textrm{L}}\left(r\right)=A_{n}^{{\rm {L}}}r^{-\alpha_{n}^{{\rm {L}}}},\\
\zeta_{n}^{\textrm{NL}}\left(r\right)=A_{n}^{{\rm {NL}}}r^{-\alpha_{n}^{{\rm {NL}}}},
\end{array} & \hspace{-0.2cm}\hspace{-0.3cm}\begin{array}{l}
\textrm{LoS:}~\textrm{Pr}_{n}^{\textrm{L}}\left(r\right)\\
\textrm{NLoS:}~1-\textrm{Pr}_{n}^{\textrm{L}}\left(r\right)
\end{array}\hspace{-0.1cm},\end{cases}\label{eq:PL_BS2UE}
\end{equation}

\noindent where $\zeta_{n}^{\textrm{L}}\left(r\right)$ and $\zeta_{n}^{\textrm{NL}}\left(r\right),n\in\left\{ 1,2,\ldots,N\right\} $
are the $n$-th piece path loss functions for the LoS transmission
and the NLoS transmission, respectively, $A_{n}^{{\rm {L}}}$ and
$A_{n}^{{\rm {NL}}}$ are the path losses at a reference distance
$r=1$ for the LoS and the NLoS cases, respectively, and $\alpha_{n}^{{\rm {L}}}$
and $\alpha_{n}^{{\rm {NL}}}$ are the path loss exponents for the
LoS and the NLoS cases, respectively. In practice, $A_{n}^{{\rm {L}}}$,
$A_{n}^{{\rm {NL}}}$, $\alpha_{n}^{{\rm {L}}}$ and $\alpha_{n}^{{\rm {NL}}}$
are constants obtainable from field tests~\cite{TR36.828,SCM_pathloss_model}.%
\begin{comment}
\noindent Besides, Note that the proposed model is general and can
be applied to several channel models that captures LoS and NLoS transmissions~{[}10-14{]}.
\end{comment}
{} Moreover, $\textrm{Pr}_{n}^{\textrm{L}}\left(r\right)$ is the $n$-th
piece LoS probability function that a transmitter and a receiver separated
by a distance $r$ have a LoS path, which is assumed to be a monotonically
decreasing function with regard to $r$.

For convenience, $\left\{ \zeta_{n}^{\textrm{L}}\left(r\right)\right\} $
and $\left\{ \zeta_{n}^{\textrm{NL}}\left(r\right)\right\} $ are
further stacked into piece-wise functions written as

\begin{singlespace}
\noindent
\begin{equation}
\zeta^{Path}\left(r\right)=\begin{cases}
\zeta_{1}^{Path}\left(r\right), & \textrm{when }0\leq r\leq d_{1}\\
\zeta_{2}^{Path}\left(r\right),\hspace{-0.3cm} & \textrm{when }d_{1}<r\leq d_{2}\\
\vdots & \vdots\\
\zeta_{N}^{Path}\left(r\right), & \textrm{when }r>d_{N-1}
\end{cases},\label{eq:general_PL_func}
\end{equation}

\end{singlespace}

\noindent where the string variable $Path$ takes the value of ``L''
and ``NL'' for the LoS and the NLoS cases, respectively.

Besides, $\left\{ \textrm{Pr}_{n}^{\textrm{L}}\left(r\right)\right\} $
is stacked into a piece-wise function as

\begin{singlespace}
\noindent
\begin{equation}
\textrm{Pr}^{\textrm{L}}\left(r\right)=\begin{cases}
\textrm{Pr}_{1}^{\textrm{L}}\left(r\right), & \textrm{when }0\leq r\leq d_{1}\\
\textrm{Pr}_{2}^{\textrm{L}}\left(r\right),\hspace{-0.3cm} & \textrm{when }d_{1}<r\leq d_{2}\\
\vdots & \vdots\\
\textrm{Pr}_{N}^{\textrm{L}}\left(r\right), & \textrm{when }r>d_{N-1}
\end{cases}.\label{eq:general_LoS_Pr}
\end{equation}

\end{singlespace}

Note that the generality and the practicality of the adopted path
loss model (\ref{eq:prop_PL_model}) have been well established in~\cite{our_work_TWC2016}.%

In this paper, we also assume a practical user association strategy
(UAS), in which each UE should be connected to the BS with the smallest
path loss (i.e., with the largest $\zeta\left(r\right)$) to the UE~\cite{Related_work_Health,our_work_TWC2016}.
Note that in our previous work~\cite{our_GC_paper_2015_HPPP} and
some existing works~\cite{Jeff2011,related_work_Jeff}, it was assumed
that each UE should be associated with the BS at the closest proximity.
Such assumption is not appropriate for the considered path loss model
in (\ref{eq:prop_PL_model}), because in practice it is possible for
a UE to connect to a BS that is not the nearest one but with a LoS
path, which is stronger that the NLoS path of the nearest one. Moreover,
we assume that each BS/UE is equipped with an isotropic antenna, and
that the multi-path fading between a BS and a UE is modeled as independently
identical distributed (i.i.d.) Rayleigh fading~\cite{related_work_Jeff,Related_work_Health,our_GC_paper_2015_HPPP,our_work_TWC2016}.
\begin{comment}
Note that a more practical Rician fading will be considered in the
simulation section to show its minor impact on our conclusions.
\end{comment}

\begin{comment}
we consider two user association strategies (UASs), which are
\begin{itemize}
\item \textit{UAS~1}: Each UE is associated with the BS with the smallest
path loss to the UE~\cite{Related_work_Health},~\cite{Book_Ming}
\item \textit{UAS~2}: Each UE is associated with the nearest BS to the
UE~\cite{Jeff's work 2011},~\cite{related_work_Jeff}
\end{itemize}
Provided that all small cell BSs transmit with the same power, UAS~1
implies a strategy that associates each UE to the BS with the strongest
signal reception strength, averaging out the multi-path fading. Using
UAS~1, it is possible for a UE to associate with a BS further way
but with a LoS path, instead of a nearest BS with a LoS path. Note
that both association strategies are widely used in the literature~\cite{Jeff's work 2011},~\cite{related_work_Jeff},~\cite{Related_work_Health},~\cite{Book_Ming}.
\end{comment}

\begin{comment}
Finally, we assume that each BS/UE is equipped with an isotropic antenna.
As a common practice in the field~{[}4-6,10,12{]}, the multi-path
fading between an arbitrary BS and an arbitrary UE is modeled as independently
identical distributed (i.i.d.) Rayleigh fading.
\end{comment}

\section{Main Results\label{sec:General-Results}}

\begin{comment}
Placeholder
\end{comment}

Using the HPPP theory, we study the performance of the SCNs %
\begin{comment}
in terms of the coverage probability and the ASE
\end{comment}
by considering the performance of a typical UE located at the origin
$o$.

\subsection{The Coverage Probability}

\begin{comment}
Placeholder
\end{comment}

First, we investigate the coverage probability that the typical UE's
SINR %
\begin{comment}
signal-to-interference-plus-noise ratio (SINR)
\end{comment}
is above a designated threshold $\gamma$:

\begin{singlespace}
\noindent
\begin{equation}
p^{\textrm{cov}}\left(\lambda,\gamma\right)=\textrm{Pr}\left[\mathrm{SINR}>\gamma\right],\label{eq:Coverage_Prob_def}
\end{equation}

\end{singlespace}

\noindent where the SINR is computed by

\noindent
\begin{equation}
\mathrm{SINR}=\frac{P\zeta\left(r\right)h}{I_{\textrm{agg}}+P_{{\rm {N}}}},\label{eq:SINR}
\end{equation}

\noindent where $h$ is the channel gain, which is modeled as an exponential
random variable (RV) with the mean of one due to our consideration
of Rayleigh fading mentioned above, $P$ and $P_{{\rm {N}}}$ are
the BS transmission power and the additive white Gaussian noise (AWGN)
power at each UE, respectively, and $I_{\textrm{agg}}$ is the cumulative
interference given by

\noindent
\begin{equation}
I_{\textrm{agg}}=\sum_{i:\,b_{i}\in\tilde{\Phi}\setminus b_{o}}P\beta_{i}g_{i},\label{eq:cumulative_interference}
\end{equation}

\noindent where $b_{o}$ is the BS serving the typical UE located
at distance $r$ from the typical UE, and $b_{i}$, $\beta_{i}$ and
$g_{i}$ are the $i$-th interfering BS, the path loss associated
with $b_{i}$ and the multi-path fading channel gain associated with
$b_{i}$, respectively. Note that in our previous works~\cite{our_GC_paper_2015_HPPP,our_work_TWC2016},
all BSs were assumed to be powered on, and thus $\Phi$ was used in
the expression of $I_{\textrm{agg}}$. Here, in (\ref{eq:cumulative_interference}),
only the activated BSs in $\tilde{\Phi}\setminus b_{o}$ inject effective
interference into the network, and thus the other BSs in idle modes
are not taken into account in the analysis of $I_{\textrm{agg}}$.

Based on the path loss model in (\ref{eq:prop_PL_model}) and the
considered UAS, we present our result of $p^{\textrm{cov}}\left(\lambda,\gamma\right)$
in Theorem~\ref{thm:p_cov_UAS1} shown on the top of the next page.
It is important to note that:

\noindent
\begin{algorithm*}[!tp]
\begin{thm}
{\small{}\label{thm:p_cov_UAS1}Considering the path loss model in
(\ref{eq:prop_PL_model}) and }the presented UAS{\small{}, the probability
of coverage $p^{{\rm {cov}}}\left(\lambda,\gamma\right)$ can be derived
as}{\small \par}

\noindent {\small{}
\begin{equation}
p^{{\rm {cov}}}\left(\lambda,\gamma\right)=\sum_{n=1}^{N}\left(T_{n}^{{\rm {L}}}+T_{n}^{{\rm {NL}}}\right),\label{eq:Theorem_1_p_cov}
\end{equation}
}{\small \par}

\noindent \vspace{-0.1cm}

\noindent {\small{}where $T_{n}^{{\rm {L}}}=\int_{d_{n-1}}^{d_{n}}{\rm {Pr}}\left[\frac{P\zeta_{n}^{{\rm {L}}}\left(r\right)h}{I_{{\rm {agg}}}+P_{{\rm {N}}}}>\gamma\right]f_{R,n}^{{\rm {L}}}\left(r\right)dr$,
$T_{n}^{{\rm {NL}}}=\int_{d_{n-1}}^{d_{n}}{\rm {Pr}}\left[\frac{P\zeta_{n}^{{\rm {NL}}}\left(r\right)h}{I_{{\rm {agg}}}+P_{{\rm {N}}}}>\gamma\right]f_{R,n}^{{\rm {NL}}}\left(r\right)dr$,
and $d_{0}$ and $d_{N}$ are defined as $0$ and $+\infty$, respectively.
Moreover, $f_{R,n}^{{\rm {L}}}\left(r\right)$ and $f_{R,n}^{{\rm {NL}}}\left(r\right)$
$\left(d_{n-1}<r\leq d_{n}\right)$, are represented by}{\small \par}

\noindent {\small{}
\begin{equation}
f_{R,n}^{{\rm {L}}}\left(r\right)=\exp\left(\hspace{-0.1cm}-\hspace{-0.1cm}\int_{0}^{r_{1}}\left(1-{\rm {Pr}}^{{\rm {L}}}\left(u\right)\right)2\pi u\lambda du\right)\exp\left(\hspace{-0.1cm}-\hspace{-0.1cm}\int_{0}^{r}{\rm {Pr}}^{{\rm {L}}}\left(u\right)2\pi u\lambda du\right){\rm {Pr}}_{n}^{{\rm {L}}}\left(r\right)2\pi r\lambda,\hspace{-0.1cm}\hspace{-0.1cm}\hspace{-0.1cm}\hspace{-0.1cm}\label{eq:geom_dis_PDF_UAS1_LoS_thm}
\end{equation}
}{\small \par}

\vspace{-0.2cm}

\noindent and

\vspace{-0.2cm}

\noindent {\small{}
\begin{equation}
\hspace{-0.1cm}\hspace{-0.1cm}\hspace{-0.1cm}\hspace{-0.1cm}\hspace{-0.1cm}\hspace{-0.1cm}f_{R,n}^{{\rm {NL}}}\left(r\right)=\exp\left(\hspace{-0.1cm}-\hspace{-0.1cm}\int_{0}^{r_{2}}{\rm {Pr}}^{{\rm {L}}}\left(u\right)2\pi u\lambda du\right)\exp\left(\hspace{-0.1cm}-\hspace{-0.1cm}\int_{0}^{r}\left(1-{\rm {Pr}}^{{\rm {L}}}\left(u\right)\right)2\pi u\lambda du\right)\left(1-{\rm {Pr}}_{n}^{{\rm {L}}}\left(r\right)\right)2\pi r\lambda,\hspace{-0.1cm}\hspace{-0.1cm}\hspace{-0.1cm}\hspace{-0.1cm}\label{eq:geom_dis_PDF_UAS1_NLoS_thm}
\end{equation}
}{\small \par}

\noindent {\small{}where $r_{1}$ and $r_{2}$ are given implicitly
by the following equations as}%
\begin{comment}
\noindent {\small{}the value of $r$ and }
\end{comment}
{\small \par}

\noindent {\small{}
\begin{equation}
r_{1}=\underset{r_{1}}{\arg}\left\{ \zeta^{{\rm {NL}}}\left(r_{1}\right)=\zeta_{n}^{{\rm {L}}}\left(r\right)\right\} ,\label{eq:def_r_1}
\end{equation}
}{\small \par}

\noindent \vspace{-0.2cm}
and

\vspace{-0.2cm}

\noindent {\small{}
\begin{equation}
r_{2}=\underset{r_{2}}{\arg}\left\{ \zeta^{{\rm {L}}}\left(r_{2}\right)=\zeta_{n}^{{\rm {NL}}}\left(r\right)\right\} .\label{eq:def_r_2}
\end{equation}
}{\small \par}

\noindent {\small{}In addition, ${\rm {Pr}}\left[\frac{P\zeta_{n}^{{\rm {L}}}\left(r\right)h}{I_{{\rm {agg}}}+P_{{\rm {N}}}}>\gamma\right]$
and ${\rm {Pr}}\left[\frac{P\zeta_{n}^{{\rm {NL}}}\left(r\right)h}{I_{{\rm {agg}}}+P_{{\rm {N}}}}>\gamma\right]$
are respectively computed by }{\small \par}

\noindent {\small{}
\begin{equation}
{\rm {Pr}}\left[\frac{P\zeta_{n}^{{\rm {L}}}\left(r\right)h}{I_{{\rm {agg}}}+P_{{\rm {N}}}}>\gamma\right]=\exp\left(-\frac{\gamma P_{{\rm {N}}}}{P\zeta_{n}^{{\rm {L}}}\left(r\right)}\right)\mathscr{L}_{I_{{\rm {agg}}}}^{{\rm {L}}}\left(\frac{\gamma}{P\zeta_{n}^{{\rm {L}}}\left(r\right)}\right),\label{eq:Pr_SINR_req_UAS1_LoS_thm}
\end{equation}
}{\small \par}

\noindent {\small{}where $\mathscr{L}_{I_{{\rm {agg}}}}^{{\rm {L}}}\left(s\right)$
is the Laplace transform of $I_{{\rm {agg}}}$ for LoS signal transmission
evaluated at $s$, which can be further written as}{\small \par}

\noindent {\small{}
\begin{equation}
\mathscr{L}_{I_{{\rm {agg}}}}^{{\rm {L}}}\left(s\right)=\exp\left(-2\pi\tilde{\lambda}\int_{r}^{+\infty}\frac{{\rm {Pr}}^{{\rm {L}}}\left(u\right)u}{1+\left(sP\zeta^{{\rm {L}}}\left(u\right)\right)^{-1}}du\right)\exp\left(-2\pi\tilde{\lambda}\int_{r_{1}}^{+\infty}\frac{\left[1-{\rm {Pr}}^{{\rm {L}}}\left(u\right)\right]u}{1+\left(sP\zeta^{{\rm {NL}}}\left(u\right)\right)^{-1}}du\right),\label{eq:laplace_term_LoS_UAS1_general_seg_thm}
\end{equation}
}{\small \par}

\noindent \vspace{-0.1cm}
{\small{}and}{\small \par}

\vspace{-0.1cm}

\noindent {\small{}
\begin{equation}
{\rm {Pr}}\left[\frac{P\zeta_{n}^{{\rm {NL}}}\left(r\right)h}{I_{{\rm {agg}}}+P_{{\rm {N}}}}>\gamma\right]=\exp\left(-\frac{\gamma P_{{\rm {N}}}}{P\zeta_{n}^{{\rm {NL}}}\left(r\right)}\right)\mathscr{L}_{I_{{\rm {agg}}}}^{{\rm {NL}}}\left(\frac{\gamma}{P\zeta_{n}^{{\rm {NL}}}\left(r\right)}\right),\label{eq:Pr_SINR_req_UAS1_NLoS_thm}
\end{equation}
}{\small \par}

\noindent {\small{}where $\mathscr{L}_{I_{{\rm {agg}}}}^{{\rm {NL}}}\left(s\right)$
is the Laplace transform of $I_{{\rm {agg}}}$ for NLoS signal transmission
evaluated at $s$, which can be further written as}{\small \par}

\noindent {\small{}
\begin{equation}
\mathscr{L}_{I_{{\rm {agg}}}}^{{\rm {NL}}}\left(s\right)=\exp\left(-2\pi\tilde{\lambda}\int_{r_{2}}^{+\infty}\frac{{\rm {Pr}}^{{\rm {L}}}\left(u\right)u}{1+\left(sP\zeta^{{\rm {L}}}\left(u\right)\right)^{-1}}du\right)\exp\left(-2\pi\tilde{\lambda}\int_{r}^{+\infty}\frac{\left[1-{\rm {Pr}}^{{\rm {L}}}\left(u\right)\right]u}{1+\left(sP\zeta^{{\rm {NL}}}\left(u\right)\right)^{-1}}du\right).\label{eq:laplace_term_NLoS_UAS1_general_seg_thm}
\end{equation}
}{\small \par}
\end{thm}
\begin{IEEEproof}
We omit the proof here due to the page limitation. We will provide
the full proof in the journal version of this paper.
\end{IEEEproof}
\end{algorithm*}

\vspace{-0.4cm}

\begin{comment}
To facilitate the following discussion, we provide some intuitive
explanation for Theorem~\ref{thm:p_cov_UAS1} so that the readers
can understand the physical meaning of each mathematical term in Theorem~\ref{thm:p_cov_UAS1}.
\end{comment}

\begin{itemize}
\item The impact of the BS selection on the coverage probability is measured
in (\ref{eq:geom_dis_PDF_UAS1_LoS_thm}) and (\ref{eq:geom_dis_PDF_UAS1_NLoS_thm}),
the expressions of which are thus based on $\lambda$.
\item The impact of $I_{\textrm{agg}}$ on the coverage probability is measured
in (\ref{eq:laplace_term_LoS_UAS1_general_seg_thm}) and (\ref{eq:laplace_term_NLoS_UAS1_general_seg_thm}).
Instead of $\lambda$, we plug $\tilde{\lambda}$ into (\ref{eq:laplace_term_LoS_UAS1_general_seg_thm})
and (\ref{eq:laplace_term_NLoS_UAS1_general_seg_thm}) because only
the activated BSs emit effective interference into the considered
SCN. %
\begin{comment}
The derivation of $\tilde{\lambda}$ will be discussed later.
\end{comment}
\end{itemize}
\begin{comment}
Placeholder
\end{comment}
\begin{comment}
The proof is similar to that in Appendix~A of our previous work~\cite{our_work_TWC2016},
but with a crucial difference that, instead of $\lambda$, $\tilde{\lambda}$
is plugged into (\ref{eq:laplace_term_LoS_UAS1_general_seg_thm})
and (\ref{eq:laplace_term_NLoS_UAS1_general_seg_thm}) because only
the activated BSs emit effective interference into the considered
SCN. The proof is omitted here for brevity and it will be fully provided
in the journal version of this work.
\end{comment}

For convenience, $\left\{ f_{R,n}^{\textrm{L}}\left(r\right)\right\} $
and $\left\{ f_{R,n}^{\textrm{NL}}\left(r\right)\right\} $ in Theorem~\ref{thm:p_cov_UAS1}
are further stacked into piece-wise functions written as

\begin{singlespace}
\noindent
\begin{equation}
f_{R}^{Path}\left(r\right)=\begin{cases}
f_{R,1}^{Path}\left(r\right), & \textrm{when }0\leq r\leq d_{1}\\
f_{R,2}^{Path}\left(r\right),\hspace{-0.3cm} & \textrm{when }d_{1}<r\leq d_{2}\\
\vdots & \vdots\\
f_{R,N}^{Path}\left(r\right), & \textrm{when }r>d_{N-1}
\end{cases}.\label{eq:general_fr_func}
\end{equation}

\end{singlespace}

\noindent Furthermore, we define the cumulative distribution function
(CDF) of $f_{R}^{Path}\left(r\right)$ as%
\begin{comment}
\noindent Then, $f_{R}^{Path}\left(r\right)$ should satisfy the following
normalization condition.
\end{comment}

\begin{singlespace}
\noindent
\begin{equation}
F_{R}^{Path}\left(r\right)=\int_{0}^{r}f_{R}^{Path}\left(u\right)du,\label{eq:general_Fr_func}
\end{equation}

\end{singlespace}

\noindent and we have $F_{R}^{\textrm{L}}\left(+\infty\right)+F_{R}^{\textrm{NL}}\left(+\infty\right)=1$.

\subsection{The Area Spectral Efficiency}

\begin{comment}
Placeholder
\end{comment}

Similar to~\cite{our_GC_paper_2015_HPPP,our_work_TWC2016}, we also
investigate the area spectral efficiency (ASE) in $\textrm{bps/Hz/km}^{2}$,
which can be defined as

\begin{singlespace}
\noindent
\begin{equation}
A^{{\rm {ASE}}}\left(\lambda,\gamma_{0}\right)=\tilde{\lambda}\int_{\gamma_{0}}^{+\infty}\log_{2}\left(1+\gamma\right)f_{\mathit{\Gamma}}\left(\lambda,\gamma\right)d\gamma,\label{eq:ASE_def}
\end{equation}

\end{singlespace}

\noindent where $\gamma_{0}$ is the minimum working SINR for the
considered SCN, and $f_{\mathit{\Gamma}}\left(\lambda,\gamma\right)$
is the probability density function (PDF) of the SINR observed at
the typical UE at a particular value of $\lambda$. It is important
to note that:
\begin{itemize}
\item Unlike~\cite{our_GC_paper_2015_HPPP,our_work_TWC2016}, $\tilde{\lambda}$
is used in the expression of $A^{{\rm {ASE}}}\left(\lambda,\gamma_{0}\right)$
because only the activated BSs make an effective contribution to the
ASE.%
\begin{comment}
The derivation of $\tilde{\lambda}$ will be discussed in the following
subsections.
\end{comment}
\item The ASE defined in this paper is different from that in~\cite{related_work_Jeff},
where a constant rate based on $\gamma_{0}$ is assumed for the typical
UE, no matter what the actual SINR value is. The definition of the
ASE in (\ref{eq:ASE_def}) is more realistic due to the SINR-dependent
rate, but it %
\begin{comment}
is more intractable to analyze, as it
\end{comment}
requires one more fold of numerical integral compared with~\cite{related_work_Jeff}.
\end{itemize}
\begin{comment}
Placeholder
\end{comment}

Based on the definition of $p^{{\rm {cov}}}\left(\lambda,\gamma\right)$
in (\ref{eq:Coverage_Prob_def}), which is the complementary cumulative
distribution function (CCDF) of SINR, $f_{\mathit{\Gamma}}\left(\lambda,\gamma\right)$
can be computed by

\begin{singlespace}
\noindent
\begin{equation}
f_{\mathit{\Gamma}}\left(\lambda,\gamma\right)=\frac{\partial\left(1-p^{{\rm {cov}}}\left(\lambda,\gamma\right)\right)}{\partial\gamma}.\label{eq:cond_SINR_PDF}
\end{equation}

\end{singlespace}

Considering the results of $p^{{\rm {cov}}}\left(\lambda,\gamma\right)$
and $A^{{\rm {ASE}}}\left(\lambda,\gamma_{0}\right)$ respectively
presented in~(\ref{eq:Coverage_Prob_def}) and~(\ref{eq:ASE_def}),
we can now analyze the two performance measures%
\begin{comment}
for the presented UAS
\end{comment}
. The key step to do so is to accurately derive the activated BS density,
i.e., $\tilde{\lambda}$.

In~\cite{dynOnOff_Huang2012}, the authors assumed that each UE should
be associated with the nearest BS%
\begin{comment}
 with the closest proximity
\end{comment}
{} and derived an approximate density of the activated BSs based on
the distribution of the HPPP Voronoi cell size. The main result in~\cite{dynOnOff_Huang2012}
is as follows,

\begin{singlespace}
\noindent
\begin{equation}
\tilde{\lambda}^{{\rm {minDis}}}\approx\lambda\left[1-\frac{1}{\left(1+\frac{\rho}{q\lambda}\right)^{q}}\right]\overset{\triangle}{=}\lambda_{0}\left(q\right),\label{eq:lambda_tilde_Huang}
\end{equation}

\end{singlespace}

\noindent where $\tilde{\lambda}^{{\rm {minDis}}}$ is the activated
BS density under the assumption that each UE is associated with the
closest BS. The empirical value of $q$ is set to 3.5~\cite{dynOnOff_Huang2012}.
The approximation was shown to be very accurate in the existing works~\cite{dynOnOff_Huang2012,Li2014_energyEff}.
However, the assumed UAS in~\cite{dynOnOff_Huang2012} is not inline
with that in our work. Therefore, we need to carefully derive $\tilde{\lambda}$
based on the considered UAS of the smallest path loss and our main
results are presented in the following subsections.

\subsection{An Upper Bound of $\tilde{\lambda}$}

\begin{comment}
Placeholder
\end{comment}

First, we propose an upper bound of $\tilde{\lambda}$ in Theorem~\ref{thm:lambda_tilde_UB}.
\begin{thm}
\label{thm:lambda_tilde_UB}Based on the path loss model in (\ref{eq:prop_PL_model})
and the presented UAS, $\tilde{\lambda}$ can be upper bounded by

\begin{singlespace}
\noindent
\begin{equation}
\tilde{\lambda}\leq\lambda\left(1-Q^{{\rm {off}}}\right)\overset{\triangle}{=}\tilde{\lambda}^{{\rm {UB}}},\label{eq:lambda_tilde_UB_thm}
\end{equation}

\end{singlespace}

\noindent \vspace{-0.2cm}
where

\vspace{-0.2cm}
\begin{equation}
Q^{{\rm {off}}}=\underset{r_{{\rm {max}}}\rightarrow+\infty}{\lim}\sum_{k=0}^{+\infty}\left\{ {\rm {Pr}}\left[w\nsim b\right]\right\} ^{k}\frac{\lambda_{\Omega}^{k}e^{-\lambda_{\Omega}}}{k!},\label{eq:Q_off_thm}
\end{equation}

\noindent \vspace{-0.1cm}
where $\lambda_{\Omega}=\rho\pi r_{{\rm {max}}}^{2}$, and

\begin{equation}
{\rm {Pr}}\left[w\nsim b\right]=\int_{0}^{r_{{\rm {max}}}}{\rm {Pr}}\left[\left.w\nsim b\right|r\right]\frac{2r}{r_{{\rm {max}}}^{2}}dr,\label{eq:Pr_w_notAss_b_thm}
\end{equation}

\noindent \vspace{-0.2cm}
and

\begin{singlespace}
\noindent \vspace{-0.2cm}
\begin{eqnarray}
\hspace{-0.2cm}\hspace{-0.2cm}\hspace{-0.2cm}{\rm {Pr}}\left[\left.w\nsim b\right|r\right]\hspace{-0.2cm} & = & \hspace{-0.2cm}\left[F_{R}^{{\rm {L}}}\left(r\right)+F_{R}^{{\rm {NL}}}\left(r_{1}\right)\right]{\rm {Pr}}^{{\rm {L}}}\left(r\right)\nonumber \\
\hspace{-0.2cm} &  & \hspace{-0.2cm}+\left[F_{R}^{{\rm {L}}}\left(r_{2}\right)+F_{R}^{{\rm {NL}}}\left(r\right)\right]\left[1-{\rm {Pr}}^{{\rm {L}}}\left(r\right)\right],\label{eq:Pr_w_notAss_b_cond_r_thm}
\end{eqnarray}

\end{singlespace}

\noindent where $F_{R}^{{\rm {L}}}\left(r\right)$ and $F_{R}^{{\rm {NL}}}\left(r\right)$
are defined in (\ref{eq:general_Fr_func}), and $r_{1}$ and $r_{2}$
are defined in (\ref{eq:def_r_1}) and (\ref{eq:def_r_2}), respectively.
\end{thm}
\begin{IEEEproof}
We omit the proof here due to the page limitation. We will provide
the full proof in the journal version of this paper.
\end{IEEEproof}

\subsection{A Lower Bound of $\tilde{\lambda}$}

\begin{comment}
Placeholder
\end{comment}

Next, we propose a lower bound of $\tilde{\lambda}$ in Theorem~\ref{thm:lambda_tilde_LB}.
\begin{thm}
\label{thm:lambda_tilde_LB}Based on the path loss model in (\ref{eq:prop_PL_model})
and the presented UAS, $\tilde{\lambda}$ can be lower bounded by

\begin{singlespace}
\noindent
\begin{equation}
\tilde{\lambda}\geq\tilde{\lambda}^{{\rm {minDis}}}\overset{\triangle}{=}\tilde{\lambda}^{{\rm {LB}}},\label{eq:lambda_tilde_LB}
\end{equation}

\end{singlespace}

\noindent where the approximate expression of $\tilde{\lambda}^{{\rm {minDis}}}$
is shown in (\ref{eq:lambda_tilde_Huang}).
\end{thm}
\begin{IEEEproof}
We omit the proof here due to the page limitation. We will provide
the full proof in the journal version of this paper.
\end{IEEEproof}

\subsection{The Proposed Approximation of $\tilde{\lambda}$}

\begin{comment}
Placeholder
\end{comment}

Considering the lower bound of $\tilde{\lambda}$ derived in Theorem~\ref{thm:lambda_tilde_LB},
and the fact that the approximate expression of $\tilde{\lambda}^{{\rm {minDis}}}$,
i.e., $\lambda_{0}\left(q\right)$, is an increasing function with
respect to $q$, we propose Proposition~\ref{prop:approx_lambda_tilde}
to approximate $\tilde{\lambda}$.
\begin{prop}
\label{prop:approx_lambda_tilde}Based on the path loss model in (\ref{eq:prop_PL_model})
and the considered UAS, we propose to approximate $\tilde{\lambda}$
by

\begin{singlespace}
\noindent
\begin{equation}
\tilde{\lambda}\approx\lambda_{0}\left(q^{*}\right),\label{eq:lambda_tilde_approx}
\end{equation}

\end{singlespace}

\noindent where $3.5\leq q^{*}\leq\underset{x}{\arg}\left\{ \lambda_{0}\left(x\right)=\tilde{\lambda}^{{\rm {UB}}}\right\} $
and $\tilde{\lambda}^{{\rm {UB}}}$ is in (\ref{eq:lambda_tilde_UB_thm}).
\end{prop}
\begin{comment}
Placeholder
\end{comment}

Note that the range of $q^{*}$ in Proposition~\ref{prop:approx_lambda_tilde}
is obtained according to the derived upper bound and lower bound of
$\tilde{\lambda}$ presented in Theorem~\ref{thm:lambda_tilde_LB}
and Theorem~\ref{thm:lambda_tilde_UB}, respectively. Apparently,
the value of $q^{*}$ depends on the specific forms of the path loss
model in (\ref{eq:general_PL_func}) and (\ref{eq:general_LoS_Pr}).
Hence, $q^{*}$ should be numerically found for specific path loss
models, which can be performed offline with the aid of simulation
and the bisection method~\cite{Bisection}.

\section{Simulation and Discussion\label{sec:Simulation-and-Discussion}}

\begin{comment}
Placeholder
\end{comment}

\vspace{-0.4cm}

\noindent \begin{center}
\begin{figure}
\noindent \begin{centering}
\vspace{-0.2cm}
\includegraphics[width=7cm]{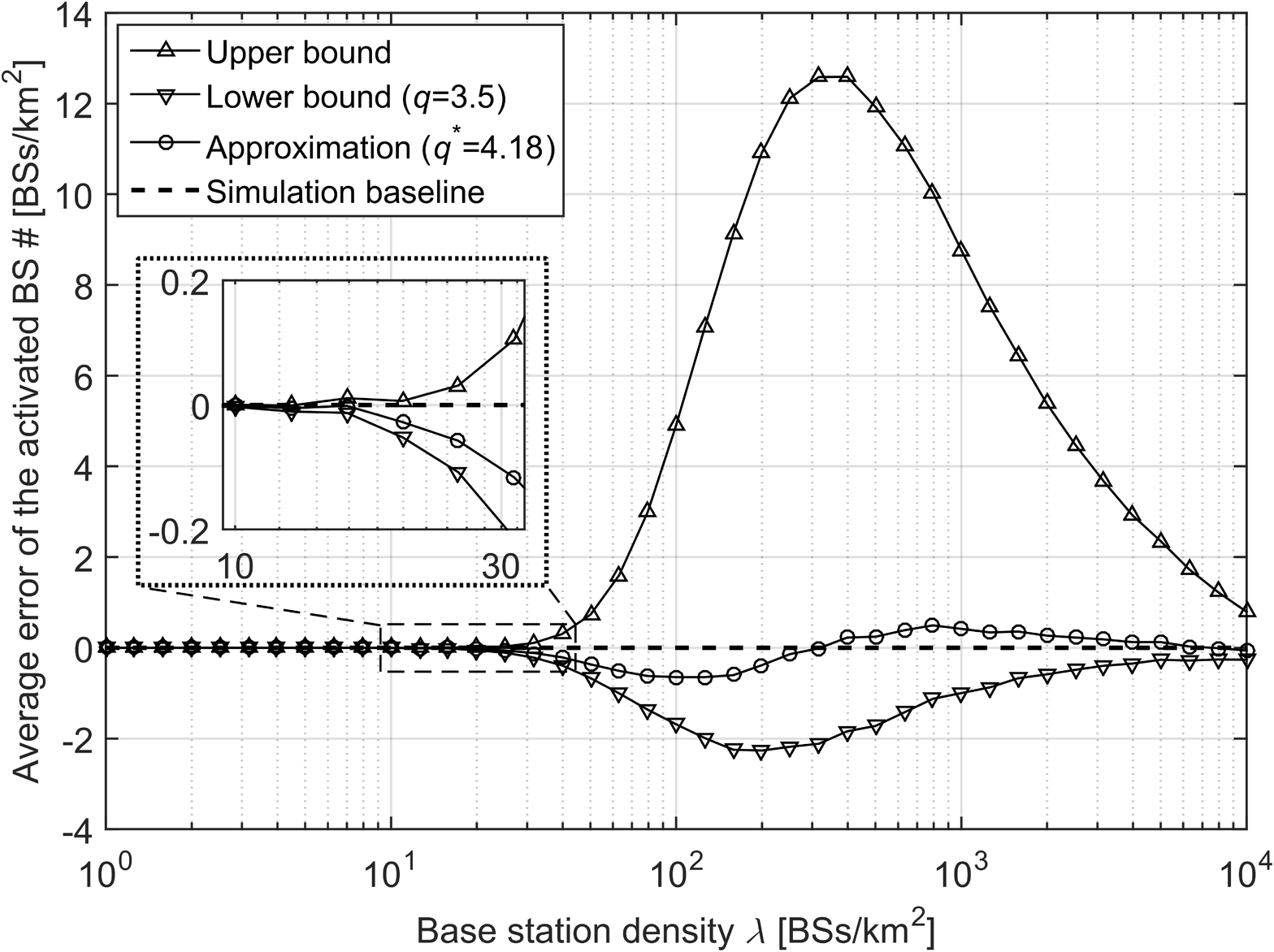}\renewcommand{\figurename}{Fig.}\caption{\label{fig:Average_error_actBSnum_UEdens300}Average error of the
activated BS number ($\rho=300\,\textrm{UEs/km}^{2}$).}
\par\end{centering}
\vspace{-0.4cm}
\end{figure}
\par\end{center}

\vspace{-0.4cm}

\begin{center}
\begin{figure}
\noindent \begin{centering}
\vspace{-0.2cm}
\includegraphics[width=7cm]{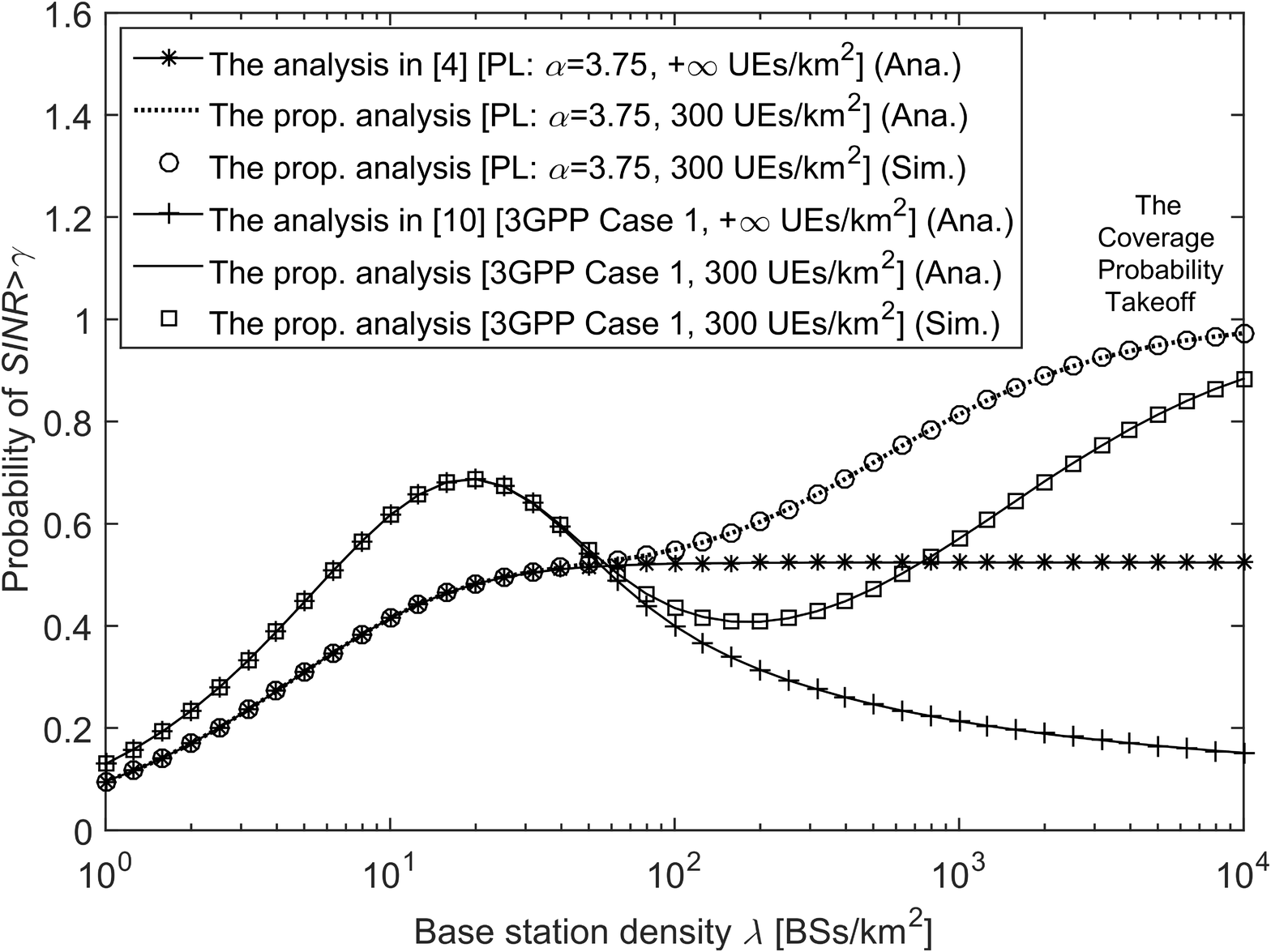}\renewcommand{\figurename}{Fig.}\caption{\label{fig:p_cov_vs_lambda_gamma0dB_UEdensity300}$p^{\textrm{cov}}\left(\lambda,\gamma\right)$
vs. $\lambda$ with $\gamma=0\,\textrm{dB}$ and $q^{*}=4.18$.}
\par\end{centering}
\vspace{-0.4cm}
\end{figure}
\par\end{center}

\vspace{-0.8cm}

In this section, we investigate the network performance and use numerical
results to validate the accuracy of our analysis.

As a special case of Theorem~\ref{thm:p_cov_UAS1}, following~\cite{our_work_TWC2016},
we consider a two-piece %
\begin{comment}
($N=2$ in (\ref{eq:prop_PL_model}))
\end{comment}
path loss and a linear LoS probability functions %
\begin{comment}
of $\textrm{Pr}^{\textrm{L}}\left(w\right)$
\end{comment}
defined by the 3GPP~\cite{TR36.828},~\cite{SCM_pathloss_model}.
Specifically, in the path loss model presented in (\ref{eq:prop_PL_model}),
we use $N=2$, $\zeta_{1}^{\textrm{L}}\left(w\right)=\zeta_{2}^{\textrm{L}}\left(w\right)=A^{{\rm {L}}}w^{-\alpha^{{\rm {L}}}}$,
and $\zeta_{1}^{\textrm{NL}}\left(w\right)=\zeta_{2}^{\textrm{NL}}\left(w\right)=A^{{\rm {NL}}}w^{-\alpha^{{\rm {NL}}}}$~\cite{TR36.828}.
Moreover, in the LoS probability model shown in (\ref{eq:general_LoS_Pr}),
we use $\textrm{Pr}_{1}^{\textrm{L}}\left(w\right)=1-\frac{w}{d_{1}}$
and $\textrm{Pr}_{2}^{\textrm{L}}\left(w\right)=0$, where $d_{1}$
is a constant~\cite{SCM_pathloss_model}. For clarity, this 3GPP
special case %
\begin{comment}
the combined path loss model of (\ref{eq:PL_BS2UE_twoslopes}) and
(\ref{eq:LoS_Prob_func_linear})
\end{comment}
is referred to as 3GPP Case~1%
\begin{comment}
 in the sequel
\end{comment}
. As justified in~\cite{our_work_TWC2016}, we use 3GPP Case~1 for
the case study because it provides tractable results for %
\begin{comment}
\noindent $\left\{ f_{R,n}^{Path}\left(r\right)\right\} $ and $\left\{ \mathscr{L}_{I_{\textrm{agg}}}^{Path}\left(s\right)\right\} $
in
\end{comment}
(\ref{eq:geom_dis_PDF_UAS1_LoS_thm})-(\ref{eq:laplace_term_NLoS_UAS1_general_seg_thm})
in Theorem~\ref{thm:p_cov_UAS1}%
\begin{comment}
by plugging (\ref{eq:PL_BS2UE_twoslopes}) and (\ref{eq:LoS_Prob_func_linear})
into (\ref{eq:geom_dis_PDF_UAS1_LoS_thm})-(\ref{eq:laplace_term_NLoS_UAS1_general_seg_thm})
\end{comment}
.%

Following~\cite{our_GC_paper_2015_HPPP,our_work_TWC2016}%
\begin{comment}
According to Tables A.1-3, A.1-4 and A.1-7 of~\cite{TR36.828} and~\cite{SCM_pathloss_model}
\end{comment}
, we adopt the following parameters for 3GPP Case~1%
\begin{comment}
An important footnote: Note that for the purpose of numerical calculation,
all the distances should be converted to km because $A^{{\rm {L}}}$
and $A^{{\rm {NL}}}$ are defined for distances in km.
\end{comment}
: $d_{1}=300$\ m, $\alpha^{{\rm {L}}}=2.09$, $\alpha^{{\rm {NL}}}=3.75$,
$A^{{\rm {L}}}=10^{-10.38}$, $A^{{\rm {NL}}}=10^{-14.54}$, $P=24$\ dBm,
$P_{{\rm {N}}}=-95$\ dBm%
\begin{comment}
 (including a noise figure of 9\ dB at the UE)
\end{comment}
. Besides, the UE density is set to $\rho=300\thinspace\textrm{UEs/km}^{2}$.%

To check the impact of different path loss models on our conclusions,
we have also investigated the results for a single-slope path loss
model that does not differentiate LoS and NLoS transmissions~\cite{Jeff2011}
where only one path loss exponent $\alpha$ is defined, the value
of which is assumed to be $\alpha=\alpha^{{\rm {NL}}}=3.75$. Note
that in this single-slope path loss model, the activated BS density
is assumed to be $\lambda_{0}\left(3.5\right)$ shown in (\ref{eq:lambda_tilde_Huang})~\cite{dynOnOff_Huang2012}.

\begin{comment}
In this section, we use numerical results to establish the accuracy
of our analysis and further study the performance of dense SCNs. According
to Tables A.1-3, A.1-4 and A.1-7 of~\cite{TR36.828} and~\cite{SCM_pathloss_model},
we adopt the following parameters for 3GPP Case~1: $d_{1}=0.3$\ km,
$\alpha^{{\rm {L}}}=2.09$, $\alpha^{{\rm {NL}}}=3.75$, $A^{{\rm {L}}}=10^{-10.38}$,
$A^{{\rm {NL}}}=10^{-14.54}$, $P=24$\ dBm, $N_{0}=-95$\ dBm (including
a noise figure of 9\ dB at the UE).
\end{comment}
{} %
\begin{comment}
In our figures, $\alpha$ is set to 3.75 $\alpha^{{\rm {L}}}$ or
$\alpha^{{\rm {NL}}}$ to show the results of the analysis from~\cite{Jeff's work 2011}.
\end{comment}
\begin{comment}
\begin{tabular}{|l|l|l|}
\hline
Parameters & Description & Value\tabularnewline
\hline
\hline
$d_{1}$ & The cut-off point in the linear LoS probability function & 300$\,$m\tabularnewline
\hline
$\alpha^{{\rm {L}}}$ & The path loss exponent for LoS & 2.09\tabularnewline
\hline
$\alpha^{{\rm {NL}}}$ & The path loss exponent for NLoS & 3.75\tabularnewline
\hline
$A^{{\rm {L}}}$ & The constant value in the path loss function for LoS & $10^{-14.54}$\tabularnewline
\hline
$A^{{\rm {NL}}}$ & The constant value in the path loss function for NLoS & $10^{-10.38}$\tabularnewline
\hline
$P$ & The BS transmission power & 24$\,$dBm\tabularnewline
\hline
$N_{0}$ & The noise power & -95$\,$dBm\tabularnewline
\hline
\end{tabular}
\end{comment}

\subsection{Discussion on the Value of $q^{*}$ for the Approximate $\tilde{\lambda}$\label{subsec:Discu_q_star}}

\begin{comment}
Placeholder
\end{comment}

Considering 3GPP Case~1 and Proposition~\ref{prop:approx_lambda_tilde},
we conduct simulations to numerically find the optimal $q^{*}$ for
the approximate $\tilde{\lambda}$. Based on the minimum mean squared
error (MMSE) criterion, we obtain $q^{*}=4.18$. Fig.~\ref{fig:Average_error_actBSnum_UEdens300}
shows the average errors of the number of activated BS for $\tilde{\lambda}^{{\rm {UB}}}$,
$\tilde{\lambda}^{{\rm {LB}}}$, and $\lambda_{0}\left(q^{*}\right)$.
Note that in Fig.~\ref{fig:Average_error_actBSnum_UEdens300} all
results are compared against the simulation results, which create
baseline results with zero errors. We can draw the following conclusions
from Fig.~\ref{fig:Average_error_actBSnum_UEdens300}:
\begin{itemize}
\item The proposed upper bound $\tilde{\lambda}^{{\rm {UB}}}$ and lower
bound $\tilde{\lambda}^{{\rm {LB}}}$ are valid compared with the
simulation results, i.e., $\tilde{\lambda}^{{\rm {UB}}}$ and $\tilde{\lambda}^{{\rm {LB}}}$
are always larger and smaller than the simulation baseline results,
respectively.
\item $\tilde{\lambda}^{{\rm {UB}}}$ is tighter than $\tilde{\lambda}^{{\rm {LB}}}$
when $\lambda$ is relatively small, e.g., $\lambda\in\left[10,30\right]\,\textrm{BSs/km}^{2}$.
\item $\tilde{\lambda}^{{\rm {LB}}}$ is much tighter than $\tilde{\lambda}^{{\rm {UB}}}$
for dense and ultra-dense SCNs, e.g., $\lambda>100\,\textrm{BSs/km}^{2}$.
\item The maximum error associated with $\lambda_{0}\left(q^{*}\right)$
is around $\pm$0.5\,$\textrm{BSs/km}^{2}$, which is smaller than
that of $\tilde{\lambda}^{{\rm {LB}}}$, e.g., an error around -2\,$\textrm{BSs/km}^{2}$
when $\lambda=100\,\textrm{BSs/km}^{2}$.
\end{itemize}

\subsection{Validation of Theorem~\ref{thm:p_cov_UAS1} on the Coverage Probability\label{subsec:Sim-p-cov-3GPP-Case-1}}

\begin{comment}
Placeholder
\end{comment}

Fig.~\ref{fig:p_cov_vs_lambda_gamma0dB_UEdensity300} shows the results
of $p^{\textrm{cov}}\left(\lambda,\gamma\right)$ with $\gamma=0\,\textrm{dB}$
and $q^{*}=4.18$ plugged into Proposition~\ref{prop:approx_lambda_tilde}.
As one can observe, our analytical results given by Theorem~\ref{thm:p_cov_UAS1}
match the simulation results very well, which validates the accuracy
of our analysis. In fact, Fig.~\ref{fig:p_cov_vs_lambda_gamma0dB_UEdensity300}
is essentially the same as Fig.~\ref{fig:comp_p_cov_4Gto5G} with
the same marker styles, except that the results for the single-slope
path loss model with $\rho=300\,\textrm{UEs/km}^{2}$ are also plotted.
Fig.~\ref{fig:p_cov_vs_lambda_gamma0dB_UEdensity300} confirms the
key observations presented in Section~\ref{sec:Introduction}:
\begin{itemize}
\item For the single-slope path loss with $\rho=+\infty\,\textrm{UEs/km}^{2}$,
\emph{}%
\begin{comment}
\emph{the BS density does NOT matter}, since
\end{comment}
the coverage probability approaches a constant for dense SCNs~\cite{Jeff2011}.
\item For 3GPP Case~1 with $\rho=+\infty\,\textrm{UEs/km}^{2}$, \emph{}%
\begin{comment}
\emph{the BS density DOES matter,} since
\end{comment}
the coverage probability decreases as $\lambda$ increases when the
network is dense enough, i.e., $\lambda>20\,\textrm{BSs/km}^{2}$,
due to the NLoS to LoS transition of interfering paths~\cite{our_work_TWC2016}.
When $\lambda$ is very large, i.e., $\lambda\geq10^{3}\,\textrm{BSs/km}^{2}$,
the coverage probability decreases at a slower pace because both the
interference and the signal powers are LoS dominated%
\begin{comment}
and thus the conclusion in~\cite{Jeff2011} is valid again
\end{comment}
~\cite{our_work_TWC2016}.
\item For both path loss models with $\rho=300\,\textrm{UEs/km}^{2}$, the
coverage probability performance continuously increases toward one
in dense SCNs due to the IMCs, i.e., \emph{the Coverage Probability
Takeoff}, as discussed in Section~\ref{sec:Introduction}.
\end{itemize}

\subsection{The Theoretical Results of the ASE\label{subsec:Sim-ASE-3GPP-Case-1}}

\begin{comment}
Placeholder
\end{comment}

In Fig.~\ref{fig:ASE_vs_lambda_gamma0dB_UEdensity300}, we show the
results of $A^{\textrm{ASE}}\left(\lambda,\gamma_{0}\right)$ with
$\gamma_{0}=0\,\textrm{dB}$ and $q^{*}=4.18$ plugged into Proposition~\ref{prop:approx_lambda_tilde}.
We can draw the following conclusions from Fig.~\ref{fig:ASE_vs_lambda_gamma0dB_UEdensity300}:
\begin{itemize}
\item For the 3GPP Case~1 with $\rho=+\infty\,\textrm{UEs/km}^{2}$, the
ASE suffers from a slow growth when $\lambda\in\left[20,200\right]\,\textrm{BSs/km}^{2}$
because of the interference transition from NLoS to LoS~\cite{our_work_TWC2016}.
After that BS density region, for both path loss models with the IMC
and $\rho=300\,\textrm{UEs/km}^{2}$, the ASEs monotonically grow
as $\lambda$ increases in dense SCNs, but with performance gaps with
regards to those of $\rho=+\infty\,\textrm{UEs/km}^{2}$ due to the
finite number of UEs.
\item The takeaway message should not be that the IMC generates an inferior
ASE in dense SCNs. Instead, as explained in Section~\ref{sec:Introduction},
since there is a finite number of UEs in the network, some BSs are
put to sleep and thus the spectrum reuse and in turn the area spectral
efficiency decreases. However, the per-UE performance should increase
as indicated by the probability of coverage shown in Fig.~\ref{fig:p_cov_vs_lambda_gamma0dB_UEdensity300}.
\item Moreover, the IMC can greatly improve the SCN energy efficiency, i.e.,
the ratio of the ASE over the total energy consumption, which will
be investigated in our future work considering practical power models
for various levels of IMCs~\cite{Tutor_smallcell}.

\end{itemize}
\noindent \begin{center}
\vspace{-0.4cm}
\begin{figure}[H]
\noindent \begin{centering}
\vspace{-0.2cm}
\includegraphics[width=7cm]{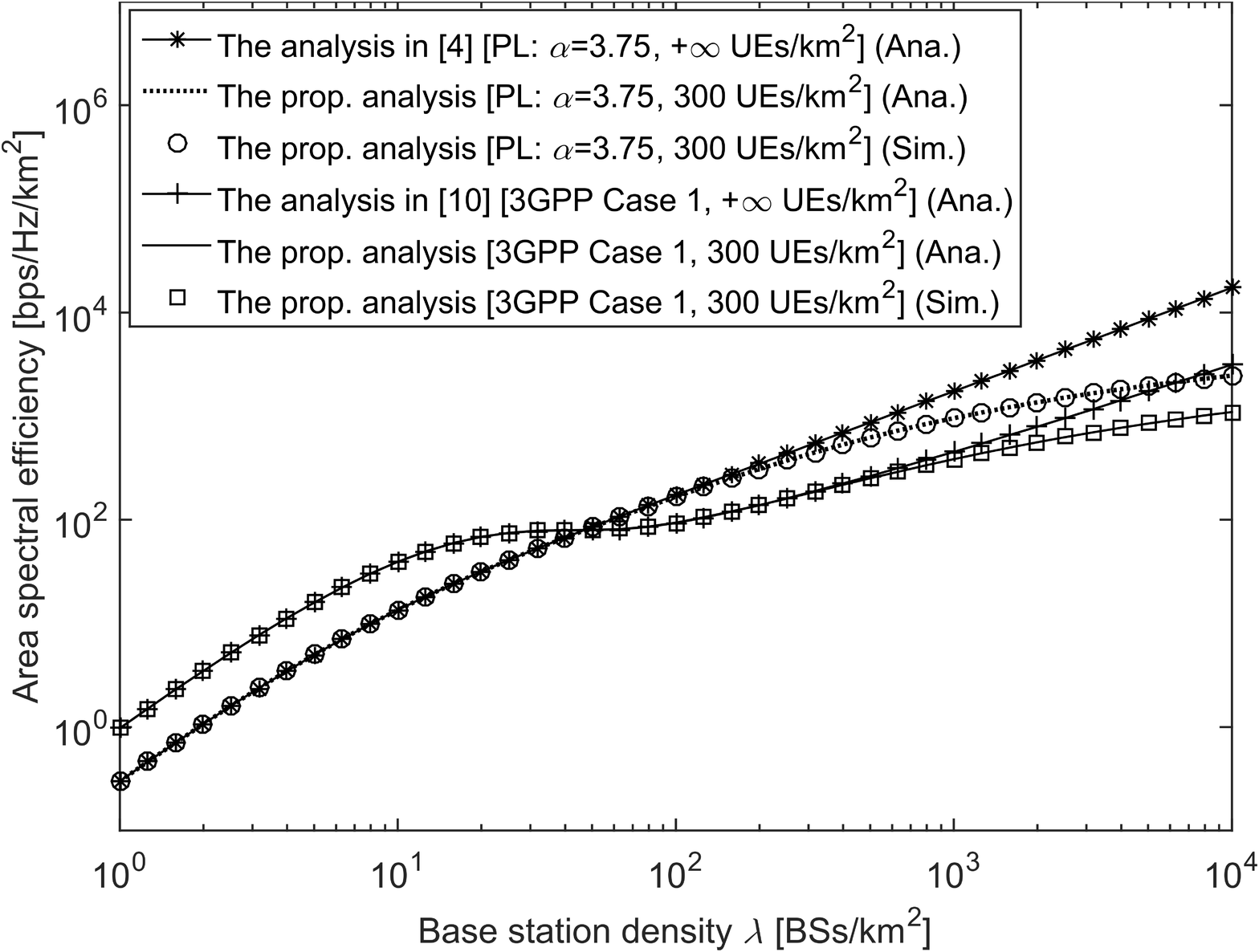}\renewcommand{\figurename}{Fig.}\caption{\label{fig:ASE_vs_lambda_gamma0dB_UEdensity300}$A^{\textrm{ASE}}\left(\lambda,\gamma_{0}\right)$
vs. $\lambda$ with $\gamma_{0}=0\,\textrm{dB}$ and $q^{*}=4.18$.}
\par\end{centering}
\vspace{-0.4cm}
\end{figure}
\par\end{center}

\vspace{-0.4cm}

\section{Conclusion\label{sec:Conclusion}}

\begin{comment}
Placeholder
\end{comment}

In this paper, we have studied the impact of the IMC on the network
performance in dense SCNs with LoS and NLoS transmissions. The impact
is significant, i.e., as the BS density surpasses the UE density,
the coverage probability will continuously increase toward one in
dense SCNs \emph{(the Coverage Probability Takeoff}), addressing the
issue caused by the NLoS to LoS transition of interfering paths.%
\begin{comment}
The performance impact of the IMC was shown to be significant in the
performance behavior of \emph{the Coverage Probability Takeoff}, i.e.,
as the BS density surpasses the UE density, the coverage probability
will continuously increase toward one in dense SCNs.
\end{comment}
{} %
\begin{comment}
As our future work, we will investigate the activated BS density for
other 3GPP path loss models, and the improvement on the energy efficiency
that the IMC brings about.
\end{comment}

\begin{comment}
Acknowlegements

Guoqiang Mao's research is supported by Australian Research Council
(ARC) Discovery projects DP110100538 and DP120102030 and Chinese National
Science Foundation project 61428102.
\end{comment}

\bibliographystyle{IEEEtran}
\bibliography{Ming_library}

\end{document}